\documentclass[journal=nalefd,manuscript=letter,layout=onecolumn]{achemso}

\usepackage[version=3]{mhchem} % Formula subscripts using \ce{}
\usepackage[T1]{fontenc}       % Use modern font encodings
\usepackage{graphicx}
\usepackage{amsmath}
\usepackage{mathrsfs}
\usepackage{amsbsy}
\usepackage{amssymb}
\usepackage{mathtools}
\usepackage{color}
\usepackage{color,soul}
\definecolor{LightGray}{gray}{0.8}
\definecolor{Orange}{rgb}{1.0, 0.31, 0.0}
\definecolor{Green}{rgb}{0.3, 1.0, 0.3}
\definecolor{Blue}{rgb}{0.75,0.75,1}

\newcommand{\fig}[1]{Fig.~\ref{#1}}
\newcommand{\figs}[1]{Figs.~\ref{#1}}

\newcommand{\bea}{\begin{eqnarray}}
\newcommand{\beal}[1]{\begin{eqnarray}\label{#1}}
\newcommand{\eea}{\end{eqnarray}}
\def\balg#1#2\ealg{\begin{align}\label{#1}#2\end{align}}
\def\balgnl#1\ealgnl{\begin{align*}#1\end{align*}}

\newcommand{\x}{{\mathbf r}}
\newcommand{\xt}{{\boldsymbol \rho}}
\newcommand{\E}{{\mathbf E}}
\renewcommand{\H}{{\mathbf H}}

\newcommand{\I}{{\mathbb I}}
\newcommand{\G}{{\mathbb G}}

\newcommand{\e}{\boldsymbol{\epsilon}}
\newcommand{\m}{\boldsymbol{\mu}}

\renewcommand{\fig}[1]{Figure~\ref{#1}}

\renewcommand{\figs}[1]{Figures~\ref{#1}}

\renewcommand{\eqref}[1]{eq~\ref{#1}}

\newcommand*{\citen}[1]{%
  \begingroup
    \romannumeral-`\x % remove space at the beginning of \setcitestyle
    \setcitestyle{numbers}%
    \cite{#1}%
  \endgroup   
}

\author{Iridanos Loulas}
\affiliation[National and Kapodistrian University of Athens]{Section of Condensed Matter Physics, National and Kapodistrian University of Athens, Panepistimioupolis, GR-157 84 Athens, Greece}
\author{Evangelos Almpanis}
\affiliation[Kapodistrian University of Athens]{Section of Condensed Matter Physics, National and Kapodistrian University of Athens, Panepistimioupolis, GR-157 84 Athens, Greece}
\alsoaffiliation[NCSR ``Demokritos'']{Institute of Nanoscience and Nanotechnology, NCSR ``Demokritos'', Patriarchou Gregoriou and Neapoleos Street, Ag. Paraskevi, GR-153 10 Athens, Greece}
\author{Minas Kouroublakis}
\affiliation[Aristotle University of Thessaloniki]{School of Informatics, Aristotle University of Thessaloniki, 54124 Thessaloniki, Greece}
\author{Kosmas L. Tsakmakidis}
\affiliation[National and Kapodistrian University of Athens]{Section of Condensed Matter Physics, National and Kapodistrian University of Athens, Panepistimioupolis, GR-157 84 Athens, Greece}
\author{Carsten Rockstuhl}
\affiliation[TFP Karlsruhe Institute of Technology]{
Institute of Theoretical Solid State Physics, Karlsruhe Institute of Technology, 76131 Karlsruhe, Germany}%
\alsoaffiliation[INT Karlsruhe Institute of Technology]{Institute of Nanotechnology, Karlsruhe Institute of Technology, 76131 Karlsruhe, Germany}

\author{Grigorios P. Zouros}
\email{zouros@mail.ntua.gr}
\affiliation[Kapodistrian University of Athens]{Section of Condensed Matter Physics, National and Kapodistrian University of Athens, Panepistimioupolis, GR-157 84 Athens, Greece}
\alsoaffiliation[National Technical University of Athens]{School of Electrical and
Computer Engineering, National Technical University of Athens, Athens 15773, Greece}
\title{Electromagnetic Multipole Theory for Two-dimensional Photonics}
\begin{document}

%\begin{tocentry}
%
%Some journals require a graphical entry for the Table of Contents.
%This should be laid out ``print ready'' so that the sizing of the
%Inside the \texttt{tocentry} environment, the font used is Helvetica
%8\,pt, as required by \emph{Journal of the American Chemical
%The surrounding frame is 9\,cm by 3.5\,cm, which is the maximum
%permitted for  \emph{Journal of the American Chemical Society}
%graphical table of content entries. The box will not resize if the
%content is too big: instead it will overflow the edge of the box.
%
%This box and the associated title will always be printed on a
%separate page at the end of the document.
%
%\end{tocentry}
\begin{abstract}
We develop a full-wave electromagnetic (EM) theory for calculating the multipole decomposition in two-dimensional (2-D) structures consisting of isolated, arbitrarily shaped, inhomogeneous, anisotropic cylinders or a collection of such. To derive the multipole decomposition, we first solve the scattering problem by expanding the scattered electric field in divergenceless cylindrical vector wave functions (CVWF) with unknown expansion coefficients that characterize the multipole response. These expansion coefficients are then expressed via contour integrals of the vectorial components of the scattered electric field evaluated via an electric field volume integral equation (EFVIE). The kernels of the EFVIE are the products of the tensorial 2-D Green's function (GF) expansion and the equivalent 2-D volumetric electric and magnetic current densities. We validate the theory using the commercial finite element solver COMSOL Multiphysics. In the validation, we compute the multipole decomposition of the fields scattered from various 2-D structures and compare the results with alternative formulations. Finally, we demonstrate the applicability of the theory to study an emerging photonics application on oligomers-based highly directional switching using active media. This analysis addresses a critical gap in current literature, where multipole theories exist primarily for three-dimensional (3-D) particles of isotropic materials. Our work enhances the understanding and utilization of the optical properties of 2-D, inhomogeneous, and anisotropic cylindrical structures, contributing to advancements in photonic and meta-optics technologies.
\end{abstract}

\noindent The study of optical particles and their interactions with light is fundamental in advancing various technological areas such as photonics and meta-optics. Multipole decomposition provides a robust framework for understanding and controlling these interactions. It involves decomposing the electromagnetic (EM) field scattered by an optical particle, or by a system of such particles, 
into a multipolar series. The individual terms in the series are distinct and correspond to dipolar, quadrupolar, octupolar, and higher-order terms. This decomposition offers profound insights into the scattering behavior and reveals the underlying physics at subwavelength scales.

The multipole analysis has been widely employed to investigate a broad spectrum of cutting-edge phenomena in three-dimensional (3-D) structures, including meta-atoms for the manipulation of EM waves \cite{Muhlig2011-bo}, design of metadimers with unique optical properties \cite{Grahn2012-fq}, Kerker scattering to control the directionality of light via multipolar interferences \cite{Fu2013-xh}, zero optical back-scattering from single nanoparticles \cite{Person2013-ij}, directed light emission thru multipolar interferences \cite{Hancu2014-re}, coupling enhancement of multipole resonances via optical beams \cite{Das2015-vc}, highly transmissive metasurfaces for polarization control\cite{Kruk2016-mg}, generalized Kerker effects in meta-optics \cite{Liu2018-tl}, selective excitation of multipolar resonances via cylindrical vector beams \cite{Manna2020-lp}, magnetic switching of Kerker scattering \cite{Zouros2020-uu}, dynamic control for invisibility to superscattering switching \cite{Zouros2021-oo}, excitation of higher-order multipolar modes via displacement resonance \cite{Tang2023-iy}, or more complex 3-D structures consisting of meta-atoms or meta-molecules such as metasurfaces and metagratings
\cite{panagiotidis2020multipolar,babicheva2021multipole,du2021optical,ra2021metagratings,babicheva2024mie}, as well as dense ensembles of plasmonic nanoparticles\cite{Dwivedi2024-eq}.

On the other hand, two-dimensional (2-D) structures, such as nanowires, long rods, oligomers, or metalattices/metagratings thereof \cite{Panta2019}, exhibit unique optical properties that can be harnessed for advanced technological applications. The multipole analysis of cylindrical structures has a plethora of applications, including efficient magnetic mirrors \cite{Liu2017-ob}, scattering invisibility in all-dielectric nanoparticles \cite{Liu2017-we}, active tuning of directional scattering of magneto-optical structures \cite{Liu2018-vp}, polarization manipulation in cylindrical metalattices \cite{Liu2018-ns}, optical beam steering control via dielectric diffractionless metasurfaces and diffractive metagratings \cite{Liu2018-oh}, tuning of toroidal dipole resonances in all-dielectric nanocylindrical metamolecules \cite{Huang2019-af}, reconfigurable metalattices via magneto-optically coated rods \cite{Liu2019-xv}, arbitrary beam steering exploiting transverse Kerker scattering \cite{Liu2021-vy}, and extreme nonreciprocal scattering in asymmetric gyrotropic cylindrical trimers \cite{Wang2022-mb}.

Complete theories for the multipole decomposition have been developed exclusively for 3-D particles of isotropic materials. These theories include the expansion of the EM field in spherical eigenvectors \cite{Grahn2012-rt}, exact expressions for source dipolar moments as well as for dipoles that radiate a definite polarization handedness \cite{Fernandez-Corbaton2015-cv}, the derivation of exact multipole moments beyond the subwavelength limit \cite{Alaee2018-gh}, an efficient multipole decomposition procedure using Lebedev and Gaussian quadrature \cite{Guo2023-ft}, and exact multipole moments for magnetic particles of arbitrary shape and size\cite{Evlyukhin2023-is}.

In this work, we provide, for the first time, a full-wave theoretical framework for the derivation of multipole decomposition in arbitrarily 2-D inhomogeneous gyrotropic structures composed either of an isolated scatterer or collections of such. This framework is undoubtedly significant for various fields of photonics, including nanoparticle engineering\cite{Zenin2020-av}, dielectric metasurfaces\cite{Zou2021-qs}, superscattering enhancement\cite{Zouros2021-oo} and Mie-tronics\cite{Koshelev2021-sg}, to name a few. In addition, being an extensive theory, it complements existing theories on multiple scattering by nanoparticle structures\cite{Berk2021-pd,Ramirez-Cuevas2022-xw} applied for photonics applications. At first, the scattered electric field $\E^{\rm sc}(\xt)$---where $\xt=(\rho,\varphi)$ is the position vector in polar coordinates---is expanded in divergenceless cylindrical vector wave functions (CVWF) with unknown expansion coefficients $A_m$, $B_m$ that determine the multipole response. $A_m$, $B_m$ are next expressed via contour integrals with integrand functions considering the vectorial components of $\E^{\rm sc}(\xt)$. In sequence, $\E^{\rm sc}(\xt)$ is calculated through a 2-D electric field volume integral equation (EFVIE) with kernels the products of the tensorial 2-D Green's function (GF) expansion $\G(\xt-\xt')$ and the equivalent 2-D volumetric electric ${\mathbf J}_{\rm eq}(\xt)$ and magnetic ${\mathbf M}_{\rm eq}(\xt)$ current densities. This scheme allows for the determination of $A_m$ and $B_m$ via 2-D volumetric integrations in the gyrotropic region of the scatterers.

We exhaustively validate our theory, first by calculating the multipole decomposition of various 2-D structures whose scattered field has been obtained from the full-wave Maxwell-solver COMSOL. We also compare our results to those obtained with the exact formulation for isotropic and gyrotropic circular cylinders \cite{pal_63}, with Mathieu functions method for elliptical cylinders \cite{Zouros2011-ay}, with the coupled-field volume integral equation-cylindrical Dini series expansion (CFVIE-CDSE) method for core-shell circular cylinders \cite{kat_zou_rou_21}, and with the method of auxiliary sources (MAS) for circular core-elliptical shell and circular dimer cylinders \cite{kou_zou_tsi_24}. In all these comparisons, an excellent agreement is observed. The main advantage that our approach provides, as compared to the above mentioned ones\cite{pal_63,Zouros2011-ay,kat_zou_rou_21,kou_zou_tsi_24} used for method verification, is that it allows the theory to be implemented in any general purpose solver, such as COMSOL, so as to perform the multipole decomposition on complicated structures composed of scatterers of non-canonical shapes with anisotropic material properties. This is a burdensome task for semi-analytical techniques, since their development requires cumbersome manipulations, while their applicability is limited, mainly, to canonical shapes, or perturbed variants.

The paper is organized as follows: In the first section, we develop the theoretical framework, then we provide various validation examples, after which there is a discussion on a photonics application on oligomers-based highly directional switching using active media, while the final section concludes the paper.

\begin{figure}[!t]
	\centering
	\includegraphics[scale=1]{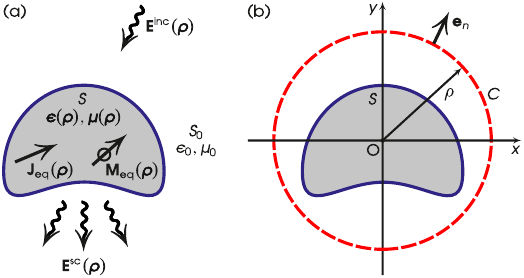}
	\caption{(a) Problem to be solved. (b) Integration path that encloses the scatterer.}
	\label{geometry}
\end{figure}

\section{Theory}\label{SOL}

\subsection{Multipole Decomposition for an Isolated Scatterer}

The configuration problem is depicted in \fig{geometry}(a). An incident electric field $\E^{\rm inc}(\xt)$ is impinging on a 2-D isolated scatterer $S$ described by an inhomogeneous, gyrotropic material with constitutive parameters
\balg{epsmu}
\!\!\!\e(\xt)=\epsilon_0
\begin{bmatrix}
\epsilon_{1r}(\xt) & i\epsilon_{2r}(\xt) & 0 \\
-i\epsilon_{2r}(\xt) & \epsilon_{1r}(\xt) & 0 \\
0 & 0 & \epsilon_{3r}(\xt)
\end{bmatrix}\text{ and }
\m(\xt)=\mu_0
\begin{bmatrix}
\mu_{1r}(\xt) & i\mu_{2r}(\xt) & 0 \\
-i\mu_{2r}(\xt) & \mu_{1r}(\xt) & 0 \\
0 & 0 & \mu_{3r}(\xt)
\end{bmatrix},
\ealg
where $\epsilon_0$ and $\mu_0$ are the free space permittivity and permeability, respectively. The background medium $S_0$ is free space. Applying the volume equivalence theorem \cite{balanis}, 2-D volumetric electric ${\mathbf J}_{\rm eq}(\xt)=i\omega[\e(\xt)-\epsilon_0\I]\E(\xt)$, $\xt\in S$, and magnetic ${\mathbf M}_{\rm eq}(\xt)=i\omega[\m(\xt)-\mu_0\I]\H(\xt)$, $\xt\in S$, current densities---where $\I$ is the unity dyadic and $\E(\xt)$ and $\H(\xt)$ the total electric and magnetic fields---are induced inside the scatterer that in turn produce the scattered electric field $\E^{\rm sc}(\xt)$ in $S_0$. The adopted time dependence is $\exp(i\omega t)$. Given the solution of the total EM field $\E(\xt)$, $\H(\xt)$ via any desirable method, the purpose is to decompose $\E^{\rm sc}(\xt)$ into a series of electric and magnetic multipoles, {\it i.e.}, to provide a theory for the multipole decomposition of any 2-D scattering problem.

For a 2-D problem, the multipoles are conveniently calculated when $\E^{\rm sc}(\xt)$ is expanded in CVWF by
\balg{eh}
\E^{\rm sc}(\xt)&=\sum_{m=-\infty}^\infty\Big[-iZ_0B_m{\mathbf M}^{(4)}_m(k_0,\xt)+A_m{\mathbf N}^{(4)}_m(k_0,\xt)\Big],
\ealg
where $A_m$, $B_m$ are the unknown expansion coefficients to be evaluated, $Z_0=(\mu_0/\epsilon_0)^{1/2}$ is free space impedance, $k_0=\omega(\epsilon_0\mu_0)^{1/2}$ is free space wavenumber, and ${\mathbf M}^{(4)}_m(k_0,\xt)$, ${\mathbf N}^{(4)}_m(k_0,\xt)$ are the CVWF of the fourth kind which, for infinitely long configurations---{\it i.e.}, $\partial/\partial z=0$---are given by \cite{kat_zou_rou_21,chew}
\balg{mn}
{\mathbf M}^{(4)}_m(k_0,\xt)&=\Big[-i\frac{m}{\rho}H_m(k_0\rho){\mathbf e}_\rho-k_0H'_m(k_0\rho){\mathbf e}_\varphi\Big]e^{-im\varphi},\notag\\
{\mathbf N}^{(4)}_m(k_0,\xt)&=k_0H_m(k_0\rho)e^{-im\varphi}{\mathbf e}_z.
\ealg
In \eqref{mn}, $H_m$ is the Hankel function of the second kind---the superscript $(2)$ is omitted for simplicity---and $H'_m$ is the derivative of $H_m$ with respect to its argument. On the right-hand side of \eqref{eh}, the term that involves the ${\mathbf M}^{(4)}_m(k_0,\xt)$ CVWF represents the TE solution, while the term that involves the ${\mathbf N}^{(4)}_m(k_0,\xt)$ CVWF represents the TM solution. For TE scattering, $m=0$ gives the magnetic dipolar (MD) response, $m=\pm1$ the electric dipolar (ED) response, and $m=\pm2$ the electric quadrupolar (EQ) response; for TM scattering, the ED, MD, and MQ contributions are given by $m=0$, $m=\pm1$ and $m=\pm2$, respectively \cite{Liu2017-we}. We aim to calculate $A_m$ and $B_m$ to determine the multipole response fully.

Next, we employ a global polar coordinate system ${\rm O}xy$ as depicted in \fig{geometry}(b). Multiplying $\E^{\rm sc}(\xt)$ with $\exp(im'\varphi)$, integrating on the circular circumference $C$ of radius $\rho$ that encloses the scatterer $S$, and utilizing the orthogonality relation of the exponential functions, $A_m$, $B_m$ are expressed via contour integrals of the components of $\E^{\rm sc}(\xt)$, {\it i.e.},
\balg{coef}
A_m&=\frac{1}{2\pi k_0\rho H_m(k_0\rho)}\int_0^{2\pi}E^{\rm sc}_z(\rho,\varphi)e^{im\varphi}\rho{\rm d}\varphi,\notag\\
B_m&=-\frac{1}{2\pi Z_0mH_m(k_0\rho)}\int_0^{2\pi}E^{\rm sc}_\rho(\rho,\varphi)e^{im\varphi}\rho{\rm d}\varphi,\notag\\
B_m&=-\frac{i}{2\pi Z_0 k_0\rho H'_m(k_0\rho)}\int_0^{2\pi}E^{\rm sc}_\varphi(\rho,\varphi)e^{im\varphi}\rho{\rm d}\varphi.
\ealg
As evident, $B_m$ is evaluated either via $E^{\rm sc}_\rho$ or $E^{\rm sc}_\varphi$. To evaluate the components of $\E^{\rm sc}(\xt)$, the 2-D scattering problem is formulated by the EFVIE \cite{kat_zou_rou_21}, namely
\balg{vie}
\E(\xt)=\E^{\rm inc}(\xt)&+(k_0^2\I+\nabla\nabla^T)\int_{\xt'\in S}g(\xt-\xt'){\mathbb X}_{\rm e}(\xt')\E(\xt'){\rm d}\xt'\notag\\
&-ik_0\sqrt{\frac{\mu_0}{\epsilon_0}}\nabla\times\int_{\xt'\in S}g(\xt-\xt'){\mathbb X}_{\rm m}(\xt')\H(\xt'){\rm d}\xt',\quad\xt\in{\mathbb R}^2.
\ealg
In \eqref{vie}, $T$ denotes transposition, $g(\xt-\xt')=-i/4H_0(k_0|\xt-\xt'|)$ is the 2-D free space GF \cite{tai_ieee}, and ${\mathbb X}_{\rm e}(\xt)=\e(\xt)/\epsilon_0-\I$ and ${\mathbb X}_{\rm m}(\xt)=\m(\xt)/\mu_0-\I$ the normalized tensorial electric and magnetic contrast function. When $\xt\in S_0$, the two 2-D volumetric integrals in \eqref{vie} represent the $\E^{\rm sc}(\xt)$; introducing the 2-D volumetric current densities ${\mathbf J}_{\rm eq}(\xt)$ and ${\mathbf M}_{\rm eq}(\xt)$, one writes
\balg{escvie}
\E^{\rm sc}(\xt)&=\frac{1}{i\omega\epsilon_0}(k_0^2\I+\nabla\nabla^T)\int_{\xt'\in S}g(\xt-\xt'){\mathbf J}_{\rm eq}(\xt'){\rm d}\xt'-\nabla\times\int_{\xt'\in S}g(\xt-\xt'){\mathbf M}_{\rm eq}(\xt'){\rm d}\xt'\notag\\
&\equiv\E^{\rm sc}_J(\xt)+\E^{\rm sc}_M(\xt),\quad\xt\in S_0.
\ealg
In what follows, we proceed separately with the evaluation of $\E^{\rm sc}_J(\xt)$ and $\E^{\rm sc}_M(\xt)$.

To evaluate $\E^{\rm sc}_J(\xt)$, the tensorial expansion $\G(\xt-\xt')$ of the 2-D GF $g(\xt-\xt')$ is used, given by eq~6 of ref~\citen{kat_zou_rou_21}. Since the 2-D volumetric integrals in \eqref{escvie} are evaluated for $\xt'\in S$, and because $\xt\in S_0$, the branch $\rho>\rho'$ is used for the $\G(\xt-\xt')$. This branch has the tensorial expansion
\balg{G}
\G(\xt-\xt')=-\frac{i}{4}\frac{1}{k_0^2}\sum_{m=-\infty}^\infty e^{-im(\varphi-\varphi')}
\begin{bmatrix}
g_{11} & g_{12} & 0 \\
g_{21} & g_{22} & 0 \\
0 & 0 & g_{33} \\
\end{bmatrix},
\ealg
where $g_{11}=\frac{m^2}{\rho\rho'}H_m(u) J_m(v)+k_0^2H'_m(u)J'_m(v)$, $g_{12}=\frac{imk_0}{\rho}H_m(u)J'_m(v)+\frac{imk_0}{\rho'}H'_m(u)J_m(v)$, $g_{21}=-\frac{imk_0}{\rho'}H'_m(u)J_m(v)-\frac{imk_0}{\rho}H_m(u)J'_m(v)$, $g_{22}=\frac{m^2}{\rho\rho'}H_m(u) J_m(v)+k_0^2H'_m(u)J'_m(v)$, $g_{33}=k_0^2H_m(u)J_m(v)$, $u=k_0\rho$, $J_m$ is the Bessel function, $J'_m$ is the derivative of $J_m$ with respect to its argument, and $v=k_0\rho'$. Substituting \eqref{G} in the $\E^{\rm sc}_J(\xt)$ term of \eqref{escvie}, expressing the $k_0^2\I+\nabla\nabla^T$ operator in cylindrical coordinates, and applying it under the integral sign on the unprimed coordinates, the $z$-component of $\E^{\rm sc}_J(\xt)$ is given by
\balg{EJz}
E^{\rm sc}_{J,z}(\xt)=-\frac{k_0^2}{4\omega\epsilon_0}\int_{\xt'\in S}\sum_{m=-\infty}^\infty e^{-im(\varphi-\varphi')}H_m(k_0\rho)J_m(k_0\rho')J_{{\rm eq},z}(\rho',\varphi'){\rm d}\xt'.
\ealg
Substituting \eqref{EJz} in the first equation of \eqref{coef} and utilizing the orthogonality relation of the exponential functions, we get
\balg{AmJ}
A_{m,J}=-\frac{Z_0}{4}\int_{\xt'\in S}e^{im\varphi'}J_m(k_0\rho')J_{{\rm eq},z}(\rho',\varphi'){\rm d}\xt'.
\ealg
The $\varphi$-component of $\E^{\rm sc}_J(\xt)$ is given by
\balgnl
&E^{\rm sc}_{J,\varphi}(\xt)=-\frac{1}{4\omega\epsilon_0}\frac{1}{k_0^2}\int_{\xt'\in S}\frac{1}{\rho}\frac{\partial}{\partial\varphi}\Big\{\frac{1}{\rho}\frac{\partial}{\partial\rho}\Big\{\rho\sum_{m=-\infty}^\infty e^{-im(\varphi-\varphi')}\notag\\
&\quad\quad\quad\quad\cdot\Big\{\Big[\frac{m^2}{\rho\rho'}H_m(k_0\rho) J_m(k_0\rho')+k_0^2H'_m(k_0\rho)J'_m(k_o\rho')\Big]J_{{\rm eq},\rho}(\rho',\varphi')\notag\\
&\quad\quad\quad\quad+\Big[\frac{imk_0}{\rho}H_m(k_0\rho)J'_m(k_0\rho')+\frac{imk_0}{\rho'}H'_m(k_0\rho)J_m(k_0\rho')\Big]J_{{\rm eq},\varphi}(\rho',\varphi')\Big\}\Big\}\Big\}{\rm d}\xt'%\notag\\
\ealgnl
\balg{EJphi}
&-\frac{1}{4\omega\epsilon_0}\int_{\xt'\in S}\sum_{m=-\infty}^\infty e^{-im(\varphi-\varphi')}\notag\\
&\quad\quad\quad\quad\cdot\Big\{\Big[\frac{-imk_0}{\rho'}H'_m(k_0\rho) J_m(k_0\rho')-\frac{imk_0}{\rho}H_m(k_0\rho)J'_m(k_o\rho')\Big]J_{{\rm eq},\rho}(\rho',\varphi')\notag\\
&\quad\quad\quad\quad+\Big[k_0^2H'_m(k_0\rho)J'_m(k_0\rho')+\frac{m^2}{\rho\rho'}H_m(k_0\rho)J_m(k_0\rho')\Big]J_{{\rm eq},\varphi}(\rho',\varphi')\Big\}{\rm d}\xt'\notag\\
&-\frac{1}{4\omega\epsilon_0}\frac{1}{k_0^2}\int_{\xt'\in S}\frac{1}{\rho}\frac{\partial}{\partial\varphi}\Big\{\frac{1}{\rho}\frac{\partial}{\partial\varphi}\sum_{m=-\infty}^\infty e^{-im(\varphi-\varphi')}\notag\\
&\quad\quad\quad\quad\cdot\Big\{\Big[\frac{-imk_0}{\rho'}H'_m(k_0\rho) J_m(k_0\rho')-\frac{imk_0}{\rho}H_m(k_0\rho)J'_m(k_o\rho')\Big]J_{{\rm eq},\rho}(\rho',\varphi')\notag\\
&\quad\quad\quad\quad+\Big[k_0^2H'_m(k_0\rho)J'_m(k_0\rho')+\frac{m^2}{\rho\rho'}H_m(k_0\rho)J_m(k_0\rho')\Big]J_{{\rm eq},\varphi}(\rho',\varphi')\Big\}\Big\}{\rm d}\xt'.
\ealg
Substituting \eqref{EJphi} in the third equation of \eqref{coef}, utilizing the orthogonality relation of the exponential functions and Bessel's differential equation $u^2 H''_m(u)+uH'_m(u)+(u^2-m^2)H_m(u)=0$---where $H''_m$ is the second derivative of $H_m$ with respect to its argument---after lengthy manipulations and cancellation of terms we get the elegant result
\balg{BmJ}
B_{m,J}=\,&\frac{m}{4}\int_{\xt'\in S}e^{im\varphi'}\frac{1}{k_0\rho'}J_m(k_0\rho')J_{{\rm eq},\rho}(\rho',\varphi'){\rm d}\xt'\notag\\
&+\frac{i}{4}\int_{\xt'\in S}e^{im\varphi'}J'_m(k_0\rho')J_{{\rm eq},\varphi}(\rho',\varphi'){\rm d}\xt'.
\ealg

To evaluate $\E^{\rm sc}_M(\xt)$, we substitute \eqref{G} in the respective term of \eqref{escvie}, express the $\nabla\times$ operator in cylindrical coordinates, and apply it under the integral sign on the unprimed coordinates. This procedure yields
\balgnl
&E^{\rm sc}_{M,z}(\xt)=-\frac{i}{4}\frac{1}{k_0^2}\int_{\xt'\in S}\frac{1}{\rho}\frac{\partial}{\partial\varphi}\Big\{\sum_{m=-\infty}^\infty e^{-im(\varphi-\varphi')}\notag\\
&\quad\quad\quad\quad\cdot\Big\{\Big[\frac{m^2}{\rho\rho'}H_m(k_0\rho) J_m(k_0\rho')+k_0^2H'_m(k_0\rho)J'_m(k_o\rho')\Big]M_{{\rm eq},\rho}(\rho',\varphi')\notag\\
&\quad\quad\quad\quad+\Big[\frac{imk_0}{\rho}H_m(k_0\rho)J'_m(k_0\rho')+\frac{imk_0}{\rho'}H'_m(k_0\rho)J_m(k_0\rho')\Big]M_{{\rm eq},\varphi}(\rho',\varphi')\Big\}\Big\}{\rm d}\xt'%\notag\\
\ealgnl
\balg{EMz}
&+\frac{i}{4}\frac{1}{k_0^2}\int_{\xt'\in S}\frac{1}{\rho}\frac{\partial}{\partial\rho}\Big\{\rho\sum_{m=-\infty}^\infty e^{-im(\varphi-\varphi')}\notag\\
&\quad\quad\quad\quad\cdot\Big\{\Big[\frac{-imk_0}{\rho'}H'_m(k_0\rho) J_m(k_0\rho')-\frac{imk_0}{\rho}H_m(k_0\rho)J'_m(k_o\rho')\Big]M_{{\rm eq},\rho}(\rho',\varphi')\notag\\
&\quad\quad\quad\quad+\Big[k_0^2H'_m(k_0\rho)J'_m(k_0\rho')+\frac{m^2}{\rho\rho'}H_m(k_0\rho)J_m(k_0\rho')\Big]M_{{\rm eq},\varphi}(\rho',\varphi')\Big\}\Big\}{\rm d}\xt'
\ealg
and
\balg{EMphi}
E^{\rm sc}_{M,\varphi}(\xt)=-\frac{i}{4}\int_{\xt'\in S}\frac{\partial}{\partial\rho}\sum_{m=-\infty}^\infty e^{-im(\varphi-\varphi')}H_m(k_0\rho)J_m(k_0\rho')M_{{\rm eq},z}(\rho',\varphi'){\rm d}\xt'.
\ealg
Employing \eqref{coef} and after lengthy manipulations we finally get
\balg{AmM}
A_{m,M}=\,&-\frac{m}{4}\int_{\xt'\in S}e^{im\varphi'}\frac{1}{k_0\rho'}J_m(k_0\rho')M_{{\rm eq},\rho}(\rho',\varphi'){\rm d}\xt'\notag\\
&-\frac{i}{4}\int_{\xt'\in S}e^{im\varphi'}J'_m(k_0\rho')M_{{\rm eq},\varphi}(\rho',\varphi'){\rm d}\xt',
\ealg
and
\balg{BmM}
B_{m,M}=-\frac{1}{4Z_0}\int_{\xt'\in S}e^{im\varphi'}J_m(k_0\rho')M_{{\rm eq},z}(\rho',\varphi'){\rm d}\xt'.
\ealg

The complete solution for the expansion coefficients $A_m$ and $B_m$ is obtained by combining \eqref{AmJ}, \eqref{BmJ}, \eqref{AmM}, and \eqref{BmM} as
\balg{AB}
A_m=A_{m,J}+A_{m,M}\quad\text{and}\quad B_m=B_{m,J}+B_{m,M}.
\ealg
Inspection of \eqref{AmJ}, \eqref{BmJ}, \eqref{AmM}, and \eqref{BmM} reveals that $B_m$ can be obtained from $A_m$, and {\itshape vice versa}, using the duality schemes $A_m\rightarrow B_m$, $J_{{\rm eq},z}\rightarrow M_{{\rm eq},z}$, $M_{{\rm eq},\rho}\rightarrow -J_{{\rm eq},\rho}$, $M_{{\rm eq},\varphi}\rightarrow -J_{{\rm eq},\varphi}$ and $B_m\rightarrow -A_m$, $M_{{\rm eq},z}\rightarrow -J_{{\rm eq},z}$, $J_{{\rm eq},\rho}\rightarrow M_{{\rm eq},\rho}$, $J_{{\rm eq},\varphi}\rightarrow M_{{\rm eq},\varphi}$, while always $\epsilon_0\leftrightarrow\mu_0$.

\begin{figure}[!t]
	\centering
	\includegraphics[scale=1]{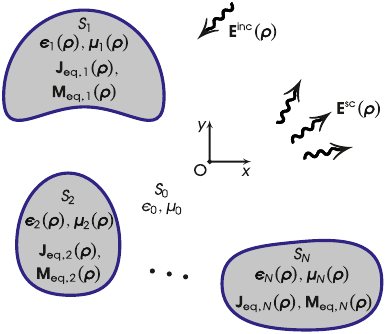}
	\caption{Multiple scattering by an arrangement of scatterers.}
	\label{multiple}
\end{figure}

\subsection{Multipole Decomposition for a Collection of Scatterers}

In this case, the configuration comprises $N$ scatterers located in free space, as depicted in \fig{multiple}. Defining a global polar coordinate system ${\rm O}xy$, each scatterer occupies domain $S_j$, $j=1,2,\ldots,N$, and is characterized by a gyrotropic material having $\e_j(\xt)$, $\m_j(\xt)$, $j=1,2,\ldots,N$. The background medium $S_0$ is free space. Upon external excitation , the 2-D volumetric current densities ${\mathbf J}_{{\rm eq},j}(\xt)$, ${\mathbf M}_{{\rm eq},j}(\xt)$, $\xt\in S_j$, $j=1,2,\ldots,N$, are induced. The collective response of all equivalent current densities produces the scattered field $\E^{\rm sc}(\xt)$ in $S_0$, {\it i.e.},
\balg{escviem}
\E^{\rm sc}(\xt)&=\sum_{j=1}^N\frac{1}{i\omega\epsilon_0}(k_0^2\I+\nabla\nabla^T)\int_{\xt'\in S_j}g(\xt-\xt'){\mathbf J}_{{\rm eq},j}(\xt'){\rm d}\xt'\notag\\
&\quad-\sum_{j=1}^N\nabla\times\int_{\xt'\in S_j}g(\xt-\xt'){\mathbf M}_{{\rm eq},j}(\xt'){\rm d}\xt'\notag\\
&\equiv\sum_{j=1}^N\E^{\rm sc}_{J,j}(\xt)+\sum_{j=1}^N\E^{\rm sc}_{M,j}(\xt),\quad\xt\in S_0.
\ealg
Therefore, the expansion coefficients $A_m$ and $B_m$ of the system, {\it i.e.}, with respect to the ${\rm O}xy$ coordinate system, are obtained by 
\balg{ABm}
A_m=\sum_{j=1}^N(A_{m,J})_j+\sum_{j=1}^N(A_{m,M})_j\quad\text{and}\quad B_m=\sum_{j=1}^N(B_{m,J})_j+\sum_{j=1}^N(B_{m,M})_j,
\ealg
where $(A_{m,J})_j$, $(A_{m,M})_j$, $(B_{m,J})_j$ and $(B_{m,M})_j$ are computed by \eqref{AmJ}, \eqref{AmM}, \eqref{BmJ} and \eqref{BmM}, respectively, for $j=1,2,\ldots,N$.

\subsection{Scattering Width, Scattering Cross-Section, Absorption Cross-Section, and Extinction Cross-Section}

By introducing the dimensionless quantities
\balg{tilde}
\tilde{A}_m\equiv A_mk_0/E_0,\quad \tilde{B}_m\equiv B_mk_0Z_0/E_0,
\ealg
where $E_0$ is the amplitude of $\E^{\rm sc}(\xt)$, the latter is written in the form
\balg{ehe0}
\E^{\rm sc}(\xt)&=E_0\sum_{m=-\infty}^\infty\Big[-i\frac{1}{k_0}\tilde{B}_m{\mathbf M}^{(4)}_m(k_0,\xt)+\frac{1}{k_0}\tilde{A}_m{\mathbf N}^{(4)}_m(k_0,\xt)\Big].
\ealg
In the far-field, $\E^{\rm sc}(\xt)\sim E_0e^{-ik_0\rho}/\sqrt{\rho}{\mathbf f}(\varphi)$, $\rho\rightarrow\infty$, where ${\mathbf f}(\varphi)$ is the scattering amplitude. Then, the scattering width $\sigma(\varphi)$, $0\leqslant\varphi<2\pi$, is given by \cite{balanis}
\balg{sw}
\sigma(\varphi)=\lim_{\rho\rightarrow\infty}2\pi\rho\frac{|{\mathbf E}^{\rm sc}(\rho,\varphi)|^2}{|{\mathbf E}^{\rm inc}(\rho,\varphi)|^2}=\frac{2\pi}{k_0}\Big[|\tilde{f}_\varphi(\varphi)|^2+|\tilde{f}_z(\varphi)|^2\Big],
\ealg
where $\tilde{f}_\varphi(\varphi)$ and $\tilde{f}_z(\varphi)$ are defined via ${\mathbf f}(\varphi)=1/\sqrt{k_0}[\tilde{f}_\varphi(\varphi){\mathbf e}_\varphi+\tilde{f}_z(\varphi){\mathbf e}_z]\equiv1/\sqrt{k_0}\tilde{{\mathbf f}}(\varphi)$, with
\balg{ftilde}
\tilde{{\mathbf f}}(\varphi)=\sqrt{\frac{2i}{\pi}}\sum_{m=-\infty}^\infty i^me^{-im\varphi}\Big(\tilde{B}_m{\mathbf e}_\varphi+\tilde{A}_m{\mathbf e}_z\Big),
\ealg
while we have assumed that $|{\mathbf E}^{\rm inc}(\rho,\varphi)|^2=1$. Equations \eqref{AmJ}, \eqref{BmJ}, \eqref{AmM}, \eqref{BmM}, \eqref{AB}, \eqref{sw}, and \eqref{ftilde} allow us to calculate the multipole contributions to the $\sigma(\varphi)$. When the TE and TM separation is valid, $\tilde{f}_\varphi(\varphi)$ corresponds to the TE scattering and $\tilde{f}_z(\varphi)$ to the TM scattering.

The full-wave scattering cross-section is given by
\balg{qscfull}
Q_{\rm sc}=\frac{1}{|{\mathbf S}^{\rm inc}(\xt)|}\oint_C{\mathbf e}_n\cdot{\mathbf S}^{\rm sc}(\xt){\rm d}\ell,
\ealg
where ${\mathbf e}_n$ is the outwards normal unit vector on the contour $C$ enclosing all scatterers---see \fig{geometry}(b) for the case of one scatterer---while ${\mathbf S}^{\rm inc}$ and ${\mathbf S}^{\rm sc}$
is the incident and scattered far-field time-averaged power flow with $|{\mathbf S}^{\rm inc}(\xt)|=E_0^2/(2Z_0)$ and $E_0=1~{\rm V/m}$. Calculating the $Q_{\rm sc}$ vs. frequency gives the spectrum of the configuration. The multipole decomposition of $Q_{\rm sc}$ is obtained via
\balg{qsc}
Q_{\rm sc}=\int_0^{2\pi}\sigma_{\rm d}(\varphi){\rm d}\varphi=\frac{4}{k_0}\sum_{m=-\infty}^\infty\Big(|\tilde{B}_m|^2+|\tilde{A}_m|^2\Big),
\ealg
where $\sigma_{\rm d}(\varphi)=|1/\sqrt{k_0}\tilde{f}_\varphi(\varphi)|^2+|1/\sqrt{k_0}\tilde{f}_z(\varphi)|^2$ is the differential scattering cross-section. The $Q_{\rm sc}$ for TE scattering is calculated solely via $\tilde{B}_m$, while for TM scattering, via $\tilde{A}_m$. For TE scattering, the $m=0$ term in \eqref{qsc} gives the MD response, the $m=\pm1$ terms the ED response, and the $m=\pm2$ terms the EQ response to the spectrum. For TM scattering, the ED, MD, and MQ responses to the spectrum are calculated by keeping the $m=0$, $m=\pm1$, and $m=\pm2$ terms, respectively, in \eqref{qsc}.

Finally, to discuss the specifics of the scattering from the point of view of energy relations, we introduce the absorption cross-section $Q_{\rm abs}$ whose full-wave evaluation is given by
\balg{qabsfull}
Q_{\rm abs}=\frac{1}{|{\mathbf S}^{\rm inc}(\xt)|}\sum_{j=1}^N\int_{\xt'\in S_j}Q_j{\rm d}\xt',
\ealg
where $Q_j$ is the power loss density of each domain $S_j$. Therefore, the full-wave extinction cross-section is given by $Q_{\rm ext}=Q_{\rm sc}+Q_{\rm abs}$\cite{bohren}, while its multipole decomposition is obtained via\cite{bohren}
\balg{qext}
Q_{\rm ext}=\frac{4}{k_0}\sum_{m=-\infty}^\infty{\rm Re}\big\{\tilde{B}_m+\tilde{A}_m\big\},
\ealg
where ${\rm Re}$ denotes the real part.

\section{Validation}\label{VAL}

Herein, we validate the developed theory by calculating the multipole decomposition of various 2-D structures and compare the results with alternative formulations. In particular, the present theory is combined with COMSOL which we use to compute the full-wave $Q_{\rm sc}$ spectrum vs. the operating frequency $f$ via \eqref{qscfull}, and its multipole contributions using \eqref{qsc}, \eqref{tilde}, \eqref{AB}, \eqref{BmM}, \eqref{AmM}, \eqref{BmJ}, and \eqref{AmJ}. Then, we compare our results with the exact formulation for isotropic and gyrotropic circular cylinders \cite{pal_63}, with Mathieu functions method for elliptical cylinders \cite{Zouros2011-ay}, with the CFVIE-CDSE for core-shell circular cylinders \cite{kat_zou_rou_21}, and with the MAS for circular core-elliptical shell and circular dimer cylinders \cite{kou_zou_tsi_24}.

\begin{figure}[!t]
	\centering
	\includegraphics[scale=1]{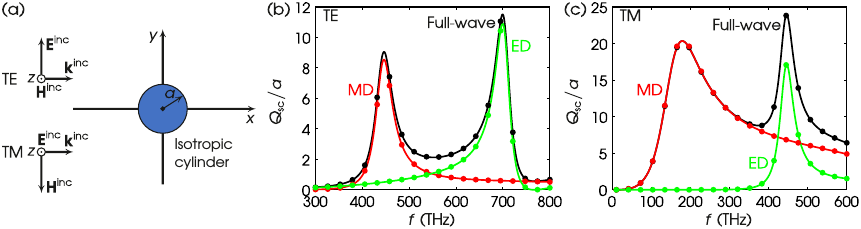}
	\caption{(a) Scattering by an isotropic circular cylinder. Values of parameters: $\epsilon_{1r}=\epsilon_{3r}=25$, $\epsilon_{2r}=0$, $\mu_{1r}=\mu_{3r}=1$, $\mu_{2r}=0$, $a=50~{\rm nm}$. (b) TE scattering. (c) TM scattering. Curves: exact; dots: this work via COMSOL; black: full-wave; red: MD; green: ED.}
	\label{cyl_iso}
\end{figure}

At first, we focus on the visible and near-IR parts of the spectrum. In \fig{cyl_iso}(a), we assume a high-index dielectric cylinder of circular cross-section. The values of the parameters are given in the figure caption. We calculate the normalized full-wave scattering cross-section, {\it i.e.}, $Q_{\rm sc}/a$, using the exact solution \cite{pal_63}, under TE illumination, as shown in \fig{cyl_iso}(b) by the black curve. The multipole decomposition is inherent in the analytical formulation, and the red and green curves show the corresponding MD and ED contributions, respectively. The calculated full-wave $Q_{\rm sc}/a$ from \eqref{qscfull}---with the aid of COMSOL---is depicted with black dots, while the corresponding multipole decomposition, using the methodology proposed here, is depicted by red---MD---and green---ED---dots. The agreement between the exact calculation and the multipole decomposition of the developed theory is apparent. The agreement is also true for TM illumination, as shown in \fig{cyl_iso}(c).

\begin{figure}[!t]
	\centering
	\includegraphics[scale=1]{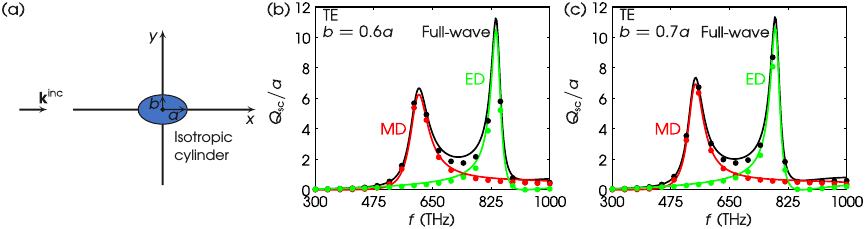}
	\caption{(a) TE scattering by an isotropic elliptical cylinder. Values of parameters: $\epsilon_{1r}=\epsilon_{3r}=25$, $\epsilon_{2r}=0$, $\mu_{1r}=\mu_{3r}=1$, $\mu_{2r}=0$, $a=50~{\rm nm}$. (b) $b=0.6a$. (c) $b=0.7a$. Curves: Mathieu functions method; dots: this work via COMSOL; black: full-wave; red: MD; green: ED.}
	\label{cyl_ell}
\end{figure}

Next, we assume a high-index dielectric cylinder with an elliptical cross-section, as shown in \fig{cyl_ell}(a). The semi-major and semi-minor axes are $a$ and $b$, respectively. We keep $a = 20~\mu{\rm m}$ and consider two different values for $b$, {\it i.e.}, $b=0.6a$ and $b=0.7a$. We assume TE illumination for both cases. In \fig{cyl_ell}(b), we depict the $Q_{\rm sc}/a$ spectrum when $b=0.6a$. The black curve corresponds to the full-wave $Q_{\rm sc}/a$ using Mathieu functions method \cite{Zouros2011-ay}, that can be decomposed to its MD---red curve---and ED---green curve---contributions. The corresponding full-wave $Q_{\rm sc}/a$ via \eqref{qscfull} is shown by black dots, while the result of the proposed multipole decomposition is shown by red---MD---and green---ED---dots. The proposed theory matches perfectly the one from \cite{Zouros2011-ay}. In \fig{cyl_ell}(c), we show the respective results for the system with $b=0.7a$.   

\begin{figure}[!t]
	\centering
	\includegraphics[scale=1]{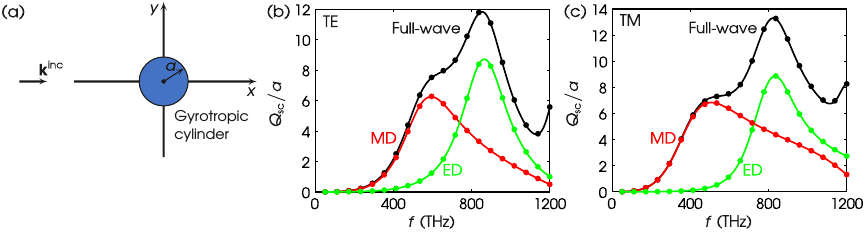}
	\caption{(a) Scattering by a gyrotropic circular cylinder. Values of parameters: $\epsilon_{1r}=4$, $\epsilon_{2r}=1$, $\epsilon_{3r}=5$, $\mu_{1r}=2$, $\mu_{2r}=0.5$, $\mu_{3r}=3$, $a=50~{\rm nm}$. (b) TE scattering. (c) TM scattering. Curves: exact; dots: this work via COMSOL; black: full-wave; red: MD; green: ED.}
	\label{cyl_gyr}
\end{figure}

In \fig{cyl_gyr}(a), we depict a cylinder with a circular cross-section consisting of gyrotropic material, {\it i.e.}, both permittivity and permeability are tensors. In \fig{cyl_gyr}(b), we plot the full-wave $Q_{\rm sc}/a$ for TE illumination, calculated using the exact solution---black curve---and the result from \eqref{qscfull}---black dots. Yet, we also plot the constituent dipolar contributions, {\it i.e.}, the MD---red curve/dots---and the ED---green curve/dots. It is apparent that the agreement between the exact solution and the proposed theory is excellent. In \fig{cyl_gyr}(c), we repeat the calculation for the TM illumination. Still, the agreement remains.

\begin{figure}[!t]
	\centering
	\includegraphics[scale=1]{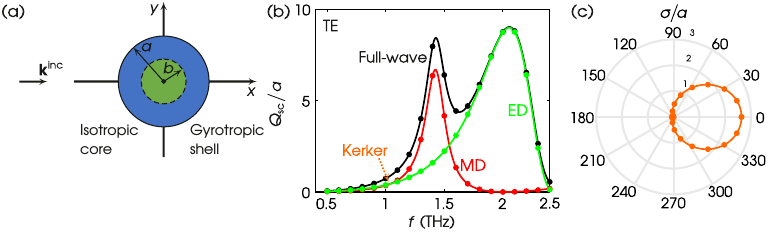}
	\caption{(a) TE scattering by a core-shell circular cylinder. Values of parameters: $\epsilon_{cr}=25$, $\mu_{cr}=1$, $\epsilon_{1r}=4$, $\epsilon_{2r}=1$, $\epsilon_{3r}=5$, $\mu_{1r}=2$, $\mu_{2r}=0.5$, $\mu_{3r}=3$, $a=20~\mu{\rm m}$, $b/a=0.75$. (b) $Q_{\rm sc}/a$ spectrum. Curves: CFVIE-CDSE; dots: this work via COMSOL; black: full-wave; red: MD; green: ED. Orange dotted arrow: location of the first MD-ED intersection at $f=1.0389$~THz. (c) $\sigma/a$ at $f=1.0389$~THz. Orange curve: CFVIE-CDSE; orange dots: this work via COMSOL.}
	\label{cyl_cs}
\end{figure}

Next, we focus our study on the THz-spectrum of frequencies. We proceed with an inhomogeneous core-shell cylinder of the circular cross-section shown in \fig{cyl_cs}(a), consisting of a high-index dielectric $b$-radius core of relative permittivity $\epsilon_{cr}$ and relative permeability $\mu_{cr}$, coated with a gyrotropic shell of outer radius $a$. The full-wave $Q_{\rm sc}/a$ for the TE illumination is calculated using the CFVIE-CDSE \cite{kat_zou_rou_21}. Results are shown by the black curve in \fig{cyl_cs}(b). In contrast, the colored curves depict the corresponding MD---red---and ED---green---contributions. The multipole decomposition is shown by dots using the respective colors. It is evident that all results agree perfectly. In addition, in \fig{cyl_cs}(c), we demonstrate the Kerker effect\cite{Evlyukhin2023-is}, i.e., zero back-scattering (first Kerker point\cite{Geffrin2012-pi}), by plotting the full-wave normalized scattering width, {\it i.e.}, $\sigma/a$, as computed by the CFVIE-CDSE and COMSOL. The Kerker effect takes place at $f=1.0389$~THz, as shown by the orange dotted arrow in \fig{cyl_cs}(b) where the MD-ED contributions intersect. Both methods in \fig{cyl_cs}(c) perfectly agree, showing zero back-scattering, thus further verifying the validity of our theory.

\begin{figure}[!t]
	\centering
	\includegraphics[scale=1]{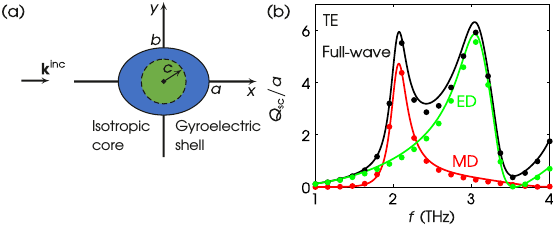}
	\caption{(a) TE scattering by a core-shell circular-elliptical cylinder. Values of parameters: $\epsilon_{cr}=25$, $\mu_{cr}=1$, $\epsilon_{1r}=4$, $\epsilon_{2r}=1$, $\epsilon_{3r}=5$, $\mu_{1r}=1$, $\mu_{2r}=0$, $\mu_{3r}=1$, $a=20~\mu{\rm m}$, $c/a=0.5$, $b/a=0.8$. (b) $Q_{\rm sc}/a$ spectrum. Curves: MAS; dots: this work via COMSOL; black: full-wave; red: MD; green: ED.}
	\label{cyl_cs_ell}
\end{figure}

In \fig{cyl_cs_ell}(a) we depict a core-shell cylinder consisting of a high-index dielectric $c$-radius circular core of relative permittivity $\epsilon_{cr}$ and relative permeability $\mu_{cr}$, coated by a gyroelectric shell of elliptical cross-section of semi-major axis $a$ and semi-minor axis $b$. Light impinges along the major axis of the ellipse in TE polarization. To calculate the full-wave $Q_{\rm sc}/a$, shown by the black curve in \fig{cyl_cs_ell}(b), we employ the MAS \cite{kou_zou_tsi_24}. The corresponding multipole decomposition via the MAS is shown by red---MD---and green---ED---curves. In addition, the black dots correspond to the full-wave $Q_{\rm sc}/a$ as computed by \eqref{qscfull}, while the red and green dots correspond to the multipole decomposition using the proposed theory. The MAS and the proposed theory behave similarly and the agreement is satisfactory.

\begin{figure}[!t]
	\centering
	\includegraphics[scale=1]{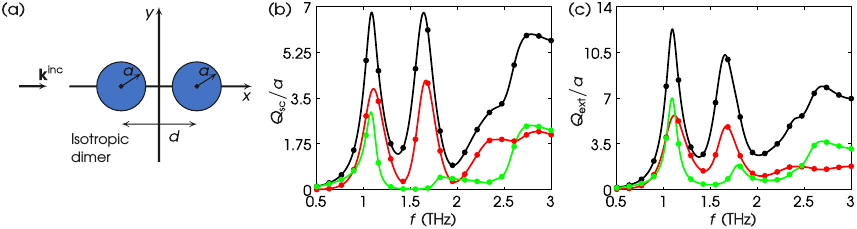}
	\caption{(a) TE scattering by a dimer. Values of parameters: $\epsilon_r=25-i2$, $\mu_r=1$, $a=20~\mu{\rm m}$, $d=60~\mu{\rm m}$. (b) $Q_{\rm sc}/a$ spectrum. Curves: MAS; dots: this work via COMSOL; black: full-wave; red: MD; green: ED. (c) Same as (b) but for $Q_{\rm ext}/a$.}
	\label{dimmer}
\end{figure}

Until now, we examined structures constituted by sole cylinders, but the theory also holds for a collection of such. As an example, we assume the dimer shown in \fig{dimmer}(a). Two identical high-index lossy dielectric cylinders of permittivity $\epsilon_{r}$ and permeability $\mu_r$ are placed with a center-to-center distance $d$. Light impinges along the $x$ axis, with TE polarization, as shown in \fig{dimmer}(a). The full-wave $Q_{\rm sc}/a$ of the dimer is calculated using the MAS, and it is depicted by the black curve in \fig{dimmer}(b). The multipole decomposition via the MAS is shown by red---MD---and green---ED---curves. The corresponding full-wave $Q_{\rm sc}/a$ via \eqref{qscfull} is shown by black dots, and the multipole decomposition via the proposed theory, by colored symbols that match perfectly the MAS solution. Since the cylinders are lossy, in \fig{dimmer}(c), we further calculate the full-wave and the multipole decomposition quantities for the $Q_{\rm ext}/a$. In particular, the full-wave $Q_{\rm ext}/a$ in COMSOL is computed by $Q_{\rm ext}/a=Q_{\rm sc}/a+Q_{\rm abs}/a$ via eqs~\ref{qscfull} and \ref{qabsfull}, while the $Q_{\rm ext}/a$ multipole decomposition from the proposed theory via \eqref{qext}. As evident, all quantities are in agreement.

The examples presented in this section confirm that the proposed theory can accurately decompose the scattered field into its magnetic and electric dipolar components, providing critical insights into the scattering mechanisms for both TE and TM illuminations. This validation suggests that the method can be reliably applied to other, more complex structures to interpret their EM scattering behavior.

\begin{figure}[!t]
	\centering
	\includegraphics[scale=1]{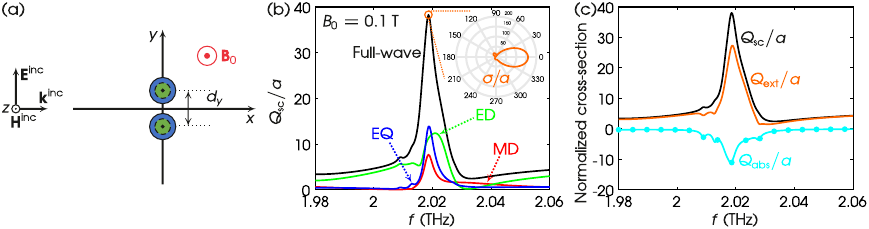}
	\caption{(a) TE scattering by a high-index core-MO shell dimer. Values of parameters: $\epsilon_{cr}=25$, $\mu_{cr}=1$, MO shell, $a=20~\mu{\rm m}$, $b/a=0.75$ and $d_y=60~\mu{\rm m}$. (b) $Q_{\rm sc}/a$ spectrum for $B_0=0.1$~T and $\alpha=-0.001$ (active). Black: full-wave; red: MD; green: ED; blue: EQ. Inset: $\sigma/a$ at $f=2.0186$~THz. (c) $Q_{\rm sc}/a$, $Q_{\rm abs}/a$, and $Q_{\rm ext}/a$ spectra for the same values of parameters as in (b). Black: full-wave $Q_{\rm sc}/a$ as computed by \eqref{qscfull}; cyan curve: full-wave $Q_{\rm abs}/a$ as computed by \eqref{qabsfull}; orange: full-wave $Q_{\rm ext}/a$ as computed by \eqref{qext}; cyan dots: $Q_{\rm ext}/a-Q_{\rm sc}/a$.
    }
	\label{app_1}
\end{figure}

\section{Application on Oligomers-based Highly Directional Switching Using Active Media}\label{APP}

The developed framework is employed to study an emerging photonics application on oligomers-based highly directional switching using active media. In \fig{app_1}(a), we employ a dimer whose elements are made of core-shell cylinders with a center-to-center distance $d_y=60~\mu{\rm m}$. Each core has radius $b=15~\mu{\rm m}$ and consists of a high-index dielectric material of relative permittivity $\epsilon_{cr}=25$ and relative permeability $\mu_{cr}=1$. Each shell has an outer radius $a=20~\mu{\rm m}$ and consists of an active magneto-optical (MO) medium which is tunable under an external magnetic flux density ${\mathbf B}_0=B_0{\mathbf e}_z$. To this end, we use indium antimonide (InSb) whose gyroelectric permittivity is given by \eqref{epsmu} with  $\epsilon_{\rm 1r}=\epsilon_\infty\{1-(\omega-iv)\omega_p^2/\{\omega[(\omega-iv)^2-\omega_c^2]\}\}$, $\epsilon_{\rm 2r}=\epsilon_\infty\{\omega_c\omega_p^2/\{\omega[(\omega-iv)^2-\omega_c^2]\}\}$ and $\epsilon_{\rm 3r}=\epsilon_\infty\{1-\omega_p^2/[\omega(\omega-iv)]\}$ \cite{Tsakmakidis2017-me}. In these definitions, $\epsilon_\infty=15.6$ accounts for interband transitions, $\omega_p=[N_ee^2/(\epsilon_0\epsilon_\infty m^\ast)]^{1/2}=4\pi\times10^{12}$~${\rm rad~s}^{-1}$ is the plasma angular frequency (with $N_e$ the electron density, $e$ the elementary charge and $m^\ast=0.0142m_e$ electron's effective mass, where $m_e$ is the electron's rest mass), $\omega_c=eB_0/m^\ast$ is the cyclotron angular frequency, and $v=\alpha\omega_p$ the damping angular frequency with $\alpha$ a dimensionless parameter. Based on the adopted time dependence $\exp(i\omega t)$, the MO material is passive (lossy) for $\alpha>0$ and active (exhibits gain) for $\alpha<0$. Therefore, by assuming the parameters of InSb, a passive material, we set negative values to the parameter $\alpha$ and make it active. In \fig{app_1}(b), we plot the full-wave $Q_{\rm sc}/a$ spectrum, as well as the multipole decomposition, when an external $B_0=0.1$~T is applied, and active shells are used in the dimer with $\alpha=-0.001$. Due to the external bias, a directional-scattering mode \cite{Zouros2021-oo} is induced at $f=2.0186$~THz, whose peak is shown by the orange circle. The multipole decomposition reveals that this mode is due to a collective contribution between the MD and EQ responses that resonate at the same frequency. The phenomenon is further enhanced by the ED response, which is almost resonant at the same frequency. Precisely at the peak, we plot the full-wave normalized scattering width, {\it i.e.}, $\sigma/a$, as computed by COMSOL. This full-wave simulation is depicted in the inset of \fig{app_1}(b), where a directional forward-scattering is shown. Since the structure in \fig{app_1}(a) incorporates an active medium, we discuss the specifics of the scattering from the point of view of energy relations, by calculating the absorption and the extinction cross-section. In \fig{app_1}(c), we plot the full-wave $Q_{\rm sc}/a$, the full-wave $Q_{\rm abs}/a$ and the full-wave $Q_{\rm ext}/a$ spectra, as computed by eqs~\ref{qscfull}, \ref{qabsfull}, and \ref{qext}, respectively. The difference $Q_{\rm ext}/a-Q_{\rm sc}/a$---cyan dots in \fig{app_1}(c)---is in perfect agreement with $Q_{\rm abs}/a$---cyan curve in \fig{app_1}(c)---proving the correctness of the calculations. The $Q_{\rm abs}/a$ is negative, indicating emission due to the active layers. Therefore, the energy scattered by the cylinder equals the energy provided by the source via the incident plane wave, plus the energy due to the emission. In the spectral window of \fig{app_1}(c), the emerging mode at $f=2.0186$~THz exhibits the maximum gain.

\begin{figure}[!t]
	\centering
	\includegraphics[scale=1]{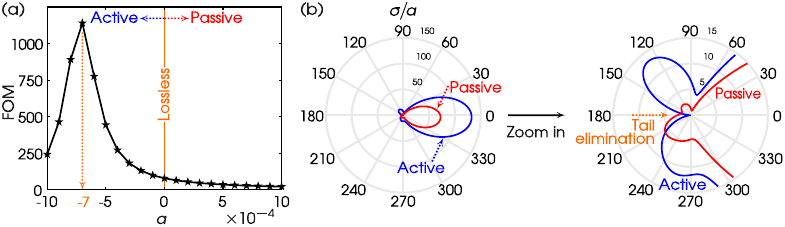}
	\caption{(a) FOM vs. $\alpha$ for the dimer of \fig{app_1}(a) when $B_0=0.1$~T and $f=2.0186$~THz. (b) $\sigma/a$ at $\alpha=\mp7\times10^{-4}$, $B_0=0.1$~T and $f=2.0186$~THz. Blue: $\alpha=-7\times10^{-4}$ (active); red: $\alpha=+7\times10^{-4}$ (passive). Zoom in: tail elimination at back-scattering.}
	\label{app_2}
\end{figure}

To reveal the advantage of using active shells in the dimer, rather than passive ones, we define the figure-of-merit (FOM) as the ratio of the forward-scattering to the back-scattering, {\it i.e.},
\balg{fom}
{\rm FOM}=\frac{\sigma(\varphi=0^\circ)}{\sigma(\varphi=180^\circ)}.
\ealg
In \fig{app_2}(a), we plot the FOM vs. various values of the dimensionless parameter $\alpha$. There exists a specific value $\alpha=-0.0007$ where the FOM is maximized. In \fig{app_2}(b), we plot the $\sigma/a$ using this specific value of $\alpha$, as well as its opposite value, {\it i.e.}, $\alpha=+0.0007$. As revealed by the zoom-in, the active shell medium ($\alpha=-0.0007$) eliminates the tail at the back-scattering direction, while the passive shell medium ($\alpha=+0.0007$) does not. Such a tail-suppression to obtain directionality is also expected in lossy structures that are illuminated using complex-frequency waves, exhibiting the so-called \emph{virtual gain}, as shown in recent publications~\cite{Kim2022-mr,proc:lou_zou_alm_tsa_24,Zouros2024-ra}.

\begin{figure}[!t]
	\centering
	\includegraphics[scale=1]{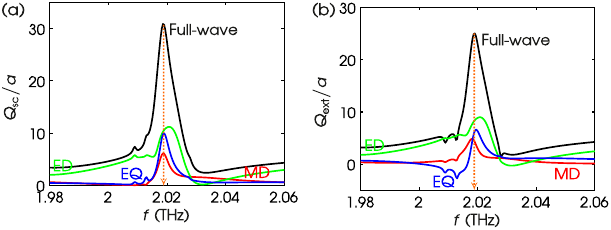}
	\caption{(a) $Q_{\rm sc}/a$ spectrum for the same values of parameters as in \fig{app_2}(b) using $\alpha=-7\times10^{-4}$. Black: full-wave; red: MD; green: ED; blue: EQ. Orange dotted arrow: location of $f=2.0186$~THz. (b) Same as (a) but for the $Q_{\rm ext}/a$ spectrum.}
	\label{app_2b}
\end{figure}

To further explain the origin of the FOM maximization that yields a back-scattering suppression at $f=2.0186$~THz when $\alpha=-0.0007$, we plot in \figs{app_2b}(a)--(b) the full-wave $Q_{\rm sc}/a$ and $Q_{\rm ext}/a$ spectra, as well as their multipole decompositions. As evident, since the structure of the dimer operates above the subwavelength regime, higher-order multipoles contribute in the construction of the emerging forward-scattering mode. Specifically at $f=2.0186$~THz, the FOM maximization---featuring the back-scattering suppression---is primarily contributed by the interference of the ED and EQ terms, with a secondary confluence from the MD term. These multipole terms interfere destructively toward the back-scattering direction of \fig{app_2}(b), producing the FOM maximization.

\begin{figure}[!t]
	\centering
	\includegraphics[scale=1]{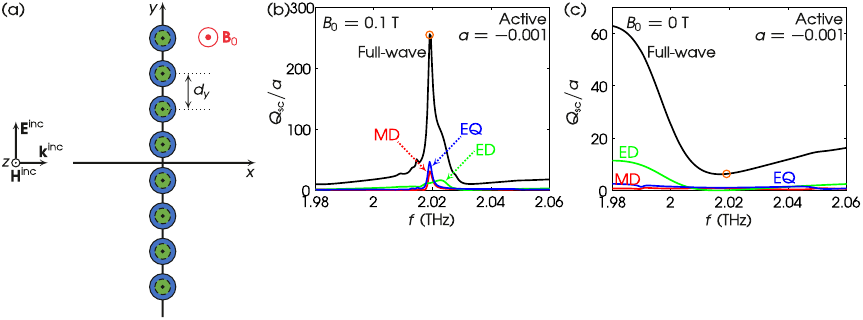}
	\caption{(a) TE scattering by a high-index core-MO shell octamer. Values of parameters: $\epsilon_{cr}=25$, $\mu_{cr}=1$, MO shell, $a=20~\mu{\rm m}$, $b/a=0.75$ and $d_y=60~\mu{\rm m}$. (b) $Q_{\rm sc}/a$ spectrum for $B_0=0.1$~T and $\alpha=-0.001$ (active). Black: full-wave; red: MD; green: ED; blue: EQ. Orange circle: Resonant mode at $f=2.0190$~THz. (c) Same as (b) but for $B_0=0$~T. Orange circle: scattering at $f=2.0190$~THz.}
	\label{app_3}
\end{figure}

To enhance the scattering and yield it highly directive, we examine in \fig{app_3}(a) a core-shell octamer having the same unit structure and values of parameters as the dimer of \fig{app_1}(a). Using active shells ($\alpha=-0.001$), in \figs{app_3}(b) and (c) we plot the $Q_{\rm sc}/a$ spectra and the multipole decompositions, when $B_0=0.1$~T and $B_0=0$~T. Due to the electrically large size of the octamer, higher-order multipole responses contribute to the spectra, however these are not depicted for graph clarity. Now, the peak (depicted by the orange circle in \fig{app_3}(b)) takes place at $f=2.0190$~THz, which is still a collective contribution of at least the MD and EQ responses. When $B_0=0$~T, a smaller scattering arises at the same frequency (depicted by the orange circle in \fig{app_3}(c)).

\begin{figure}[!t]
	\centering
	\includegraphics[scale=1]{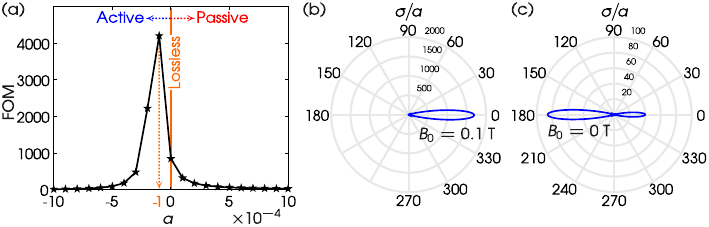}
	\caption{(a) FOM vs. $\alpha$ for the octamer of \fig{app_3}(a) when $B_0=0.1$~T and $f=2.0190$~THz. (b) $\sigma/a$ at $\alpha=-1\times10^{-4}$ (active), $B_0=0.1$~T and $f=2.0190$~THz.
    (c) Same as (b) but for $B_0=0$~T.}
	\label{app_4}
\end{figure}

Finally, to optimize the forward-scattering of the octamer's resonant mode, in \fig{app_4}(a), we show that for $\alpha=-0.0001$, the FOM becomes maximum. With this value of $\alpha$, in \fig{app_4}(b), we plot the $\sigma/a$ when the magnetic bias is at on-state. A highly directive lobe is revealed at the forward-scattering with tail elimination at the back-scattering. When the magnetic bias is at off-state, \fig{app_4}(c) shows a back-scattering with a remaining forward-scattering. Nevertheless, this on-off state suggests that a core-shell oligomer with active medium shells can operate as a highly directional active switching device in contemporary photonics.

\section{Conclusion}\label{CON}

In this work, we developed and validated a comprehensive EM multipole decomposition framework for 2-D structures, with a specific focus on inhomogeneous and anisotropic cylindrical scatterers. Our method leverages the expansion of scattered fields using divergenceless CVWF and employs 2-D volumetric integrals to express the unknown expansion coefficients. Our results on multipole decomposition using finite element simulations were validated by comparing our results with analytical and numerical methods that provide inherently the multipole decomposition. Furthermore, we demonstrated the applicability of the developed framework by analyzing photonic applications on oligomers-based highly directional-scattering switching using active MO media. Our findings reveal that the collective contributions of magnetic and electric multipole responses can be harnessed to achieve highly directional forward-scattering, a promising avenue for future photonic devices.

\begin{acknowledgement}
The authors would like to thank Ivan Fernandez-Corbaton, Karlsruhe Institute of Technology, Germany, for fruitful discussions.

I.L., E.A., K.L.T., and G.P.Z. acknowledge support for this research by the General Secretariat for Research and Technology (GSRT) and the Hellenic
Foundation for Research and Innovation (HFRI) under Grant No. 4509. K.L.T.'s part was also carried out within the framework of the National Recovery and Resilience Plan Greece 2.0, funded by the European Union—Next Generation EU (Implementation body: HFRI) under Grant No. 16909.
C.R. acknowledges support by the Deutsche Forschungsgemeinschaft (DFG, German Research Foundation) under Germany’s Excellence Strategy via the Excellence Cluster 3D Matter Made to Order (EXC-2082/1-390761711) and from the Carl Zeiss Foundation via the CZF- Focus@HEiKA Program.
\end{acknowledgement}

%\begin{suppinfo}

%A listing of the contents of each file supplied as Supporting Information
%\end{suppinfo}


\providecommand{\latin}[1]{#1}
\makeatletter
\providecommand{\doi}
  {\begingroup\let\do\@makeother\dospecials
  \catcode`\{=1 \catcode`\}=2 \doi@aux}
\providecommand{\doi@aux}[1]{\endgroup\texttt{#1}}
\makeatother
\providecommand*\mcitethebibliography{\thebibliography}
\csname @ifundefined\endcsname{endmcitethebibliography}
  {\let\endmcitethebibliography\endthebibliography}{}

\end{document}